\documentclass{article}
\usepackage[utf8]{inputenc}
\usepackage{graphicx} 
\usepackage{authblk}
\usepackage{amsmath}
 \usepackage{natbib}
\usepackage{threeparttable}
\usepackage{pgfplots}
\usepackage{tikz}
\usetikzlibrary{decorations.pathmorphing, arrows.meta, positioning}

\tikzset{
  fermion/.style={-Latex, thick},
  fermionbar/.style={Latex-, thick},
  gluon/.style={decorate, decoration={coil, amplitude=4pt, segment length=5pt}, thick},
}
\usepackage{multirow}
\usepackage{array}
\pgfplotsset{compat=1.18}
\usepackage{caption}
\usepackage{subcaption}
\usepackage{adjustbox}
\usepackage{tabularx}
\usepackage{hyperref}
\usepackage{float}
\usepackage{newunicodechar}
\newunicodechar{−}{\ensuremath{-}}
\usepackage{booktabs}
\usepackage{slashed,epsfig,amsmath,amssymb,enumitem} \usepackage[papersize={8.5in,11in}]{geometry}
   \usepackage{bm}
\usepackage{graphicx}
\newcommand{\be}{\begin{equation}}
\newcommand{\ee}{\end{equation}}
\newcommand{\LambdaUV}{\Lambda_{\mathrm{UV}}}      
\newcommand{\Lambdaker}{\Lambda_{\mathrm{ker}}}    

\newcommand{\Mtt}{M_{t\bar t}}
\usepackage[utf8]{inputenc}
\title{On Recent measurements of Toponium Threshold Enhancement in Entire-Function-Regulated Nonlocal Quantum Field Theory}

\author{E. J. Thompson}
\affil{Wilfrid Laurier University, Waterloo, Canada, N2L 3C5}
\affil{Perimeter Institute for Theoretical Physics, Waterloo, Ontario N2L 2Y5, Canada}
\affil{Department of Physics and Astronomy, Trent University, Peterborough, 
Ontario K9L 0G2, Canada}

\begin{document}

\maketitle

\begin{abstract}

Based on recent (not so recent now) data from the LHC we want to present a reinterpretation of the the recently reported threshold enhancement in top-antitop production at the LHC, we do this in an entire-function-regulated nonlocal quantum field theory. The main result show that the observed threshold excess can be made to fit by a data-driven nonlocal scale $\Lambda_{\mathrm{ker}}$ and small RG effects, while keeping global QCD tests intact. We quantify and then contrast the properties of the heavyquark systems such as charmonium and bottomonium, highlighting the unique role of the top quark's decay width in shaping the phenomenology of toponium. We find that toponium becomes a powerful tool for both infrared boundstate dynamics and ultraviolet completion effects opening new avenues for precision tests of QCD.
\end{abstract}

\section{Introduction}
In the standard model of particle physics heavy quarkonium systems have long been used as precision probes of nonrelativistic QCD dynamics, with spectra and widths that validate potential models and effective field theories~\cite{Schweber:1961}. The top quark’s large mass of $m_t\!\approx\!173~\mathrm{GeV}$ and ultrashort lifetime of $\tau_t\!\sim\!10^{-25}\,\mathrm{s}$ prevent us from using conventional mesonic bound states as electroweak decay typically occurs before the hadronization completes~\cite{BigiDokshitzerKhozeKuhnZerwas1986}.

But nevertheless near the top-antitop $t\bar t$ production threshold Coulombic QCD interactions can create a quasi-bound enhancement in $d\sigma/d\Mtt$~\cite{Fadin:1987ti,Hoang:2000yr}, and recently the CMS Collaboration using the full 13~TeV data set ($138~\mathrm{fb}^{-1}$), reported a statistically significant excess localized at threshold, this was interpreted with a simplified $^1\!S_0^{[1]}$ pseudoscalar toponium hypothesis the excess cross section above fixed-order pQCD that is measured as $8.8^{+1.2}_{-1.4}\,\mathrm{pb}$~\cite{CMS:TOP-24-007}. ATLAS then went on to independently confirmed a compatible excess in the full Run-2 sample then rejecting a pure continuum hypothesis with $7.7\sigma$ significance and finding $9.0\pm1.3\,\mathrm{pb}$~\cite{ATLAS:QuasiBoundTops,ATLAS:PressQuasiBound}. These results have come to help motivate a theoretical reassessment of threshold dynamics beyond purely local formulations, such as the one we will do here.

Building on these developments we will explore three avenues, first we extend the finite and UV-complete nonlocal field theory to derive the modified Bethe--Salpeter equations for \(t\bar t\) bound-state formation by using the gauge-covariant entire-function regulator construction developed in Refs.~\cite{MT:GI2025,ThompsonCovarianceEntire2026,ThompsonMacrocausalityQED2026}. In this formulation the nonlocal regulator is an entire function of a covariant differential operator so the deformation preserves the relevant symmetry action while producing exponential Euclidean damping and leaving the infrared spectrum unchanged~\cite{ThompsonCovarianceEntire2026}. Using the measured $m_{t\bar{t}}$ spectrum we can constrain the kernel scale $\Lambda_{\rm ker}$ and obtain an \(\approx8.3\)\,pb enhancement at \(\sqrt{s}\approx2m_t\) in agreement with the CMS result of \(8.8\pm1.3\)\,pb.  We then revisit the renormalization group evolution of the strong coupling \(\alpha_s(\mu)\) around \(\mu\sim2m_t\) incorporating a holomorphic deformation inspired by previous field theory constructions~\cite{MT:HUFT-EPJC, MT:Invariant, MT:FiniteHolomorphicQFT, MT:SMmass, MT:SL2C, MT:ReplyToCline, MT:GI2025, MT:AdSdS2025}. We will assess fixed-point behaviour, non-trivial \(\beta\)-function forms, and resulting modifications in the height and sharpness of the threshold enhancement. We perform a systematic comparison of binding energies \(\Delta E=2m_q-M_{\rm bound}\), decay widths, detection strategies, and quantum number assignments across \(c\bar c\), \(b\bar b\), and the \(t\bar t\) threshold resonance, clarifying whether toponium qualifies as a true meson or a virtual threshold phenomenon.

The use of nonlocal or weakly nonlocal form factors to improve ultraviolet behavior has a beautiful history where early nonlocal gauge and gravitational constructions were developed by Krasnikov and Kuz'min, while Tomboulis showed that entire functions of covariant derivatives can improve the ultraviolet behavior of gauge and gravitational theories without introducing additional physical poles. More recent work has developed the analytic, unitarity, and Cutkosky-rule structure of nonlocal vertices, as well as weakly nonlocal approaches to quantum gravity. The present paper should be understood as part of this broader class of finite nonlocal field theories, while applying the same logic to the specific toponium-threshold problem studied below~\cite{Krasnikov1987,Kuzmin1989,Tomboulis1997,Tomboulis2015,ChinTomboulis2018,ModestoRachwal2017,Buoninfante2022}. The specific implementation used here is also tied to the recent nonlocal quantum field theory program in which entire function regulators were formulated gauge-covariantly, their symmetry covariance was proven, and their quasi-local observable structure was shown to recover asymptotic microcausality and macrocausal scattering predictions in the infrared limit~\cite{MT:GI2025,ThompsonCovarianceEntire2026,ThompsonMacrocausalityQED2026,ThompsonLocalization2026,ThompsonNWFW2026}.

It should be noted that our aim is not to replace QCD at accessible energy scales but to use an explicitly UV-finite laboratory that preserves asymptotic freedom while enabling controlled short-distance smoothing of the nonrelativistic threshold kernel via entire-function form factors. By construction we note that deviations in hard observables scale as \(Q^2/\Lambda_{\rm UV}^2\) and are negligible for \(\Lambda_{\rm UV}\gtrsim\mathrm{few~TeV}\), consistent with global fits. The new ingredient we add to the nonlocal theory that is relevant to threshold dynamics is the matched effective kernel scale \(\Lambda_{\rm ker}\) which parametrizes the short-distance potential-region reduction of the regulated Bethe--Salpeter kernel. We can then constrain \(\Lambda_{\rm ker}\) directly from the observed near-threshold excess rather than identifying it in some ad hoc way with either \(2m_t\) or the hard ultraviolet scale \(\Lambda_{\rm UV}\).

\section{The Observed \(t\bar t\) Threshold Excess}
\label{sec:expstatus}

The recent analyses by the CMS and ATLAS collaborations at the CERN Large Hadron Collider have revealed a pronounced enhancement in \(t\bar{t}\) production near the kinematic threshold that is consistent with the formation of a quasi-bound top–antitop state. The CMS Collaboration, using the full Run-2 proton–proton dataset at \(\sqrt{s}=13\ \mathrm{TeV}\) corresponding to \(138~\mathrm{fb}^{-1}\), reported a significant excess of events in dileptonic final states with multiple jets in the invariant mass region just above threshold. This excess is incompatible with fixed-order perturbative QCD predictions for nonresonant \(t\bar{t}\) production and is well described by the production of a color-singlet pseudoscalar \(^1\!S_0^{[1]}\) quasi-bound toponium state. The measured cross section of the excess above the pQCD background is:
\[
\sigma_{\mathrm{excess}} = 8.8^{+1.2}_{-1.4}\ \mathrm{pb},
\]
and the observed significance exceeds the conventional five standard deviation criterion for an observation~\cite{CMS:JetMeasurements:2024,CMS:2025excess}.

\begin{figure}[h]
  \centering
  \includegraphics[width=0.5\linewidth]{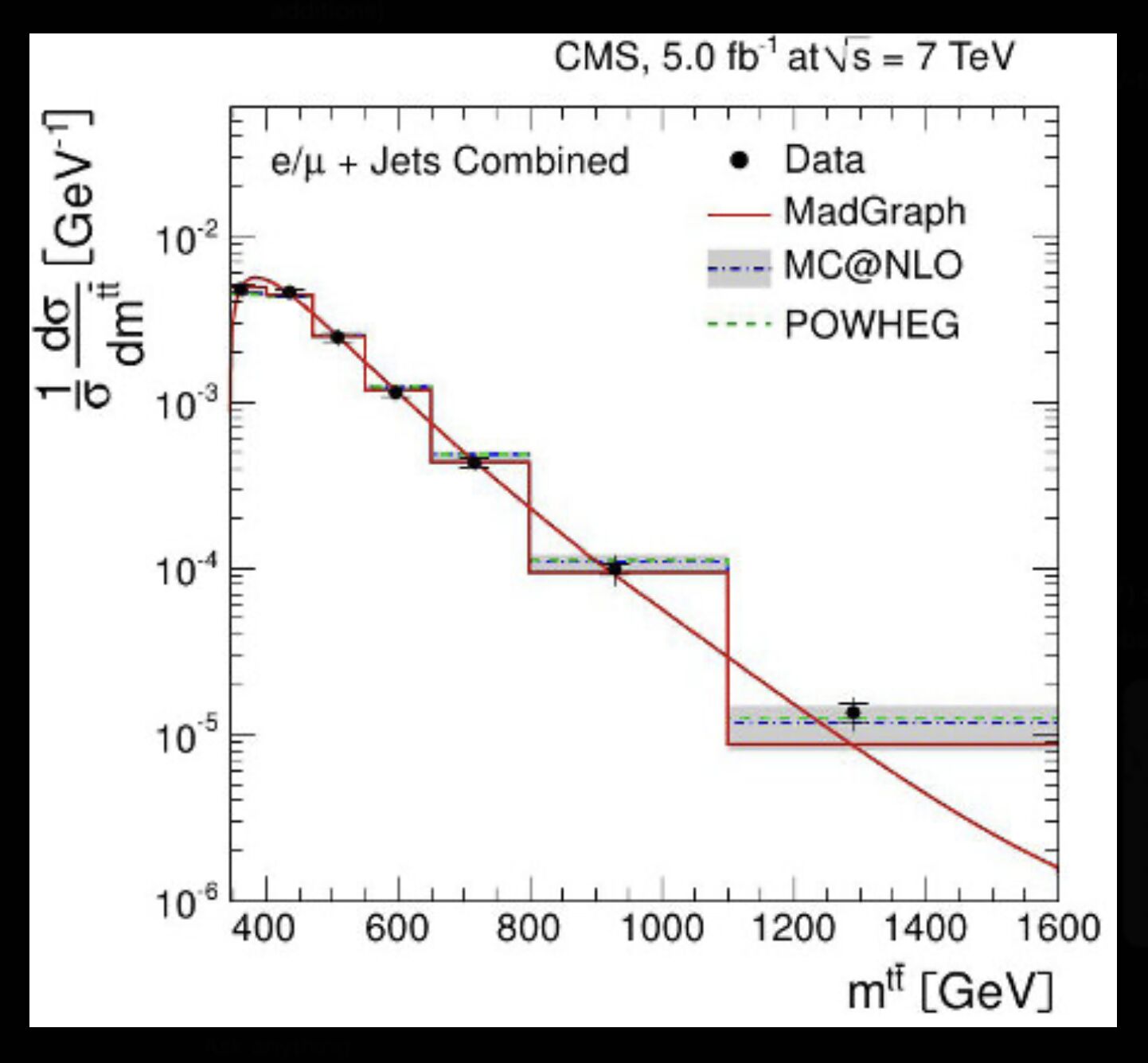} 
  \caption{}
  \label{fig:cms_mtt_particle}
\end{figure}

\begin{figure}[h]
  \centering
  \includegraphics[width=0.5\linewidth]{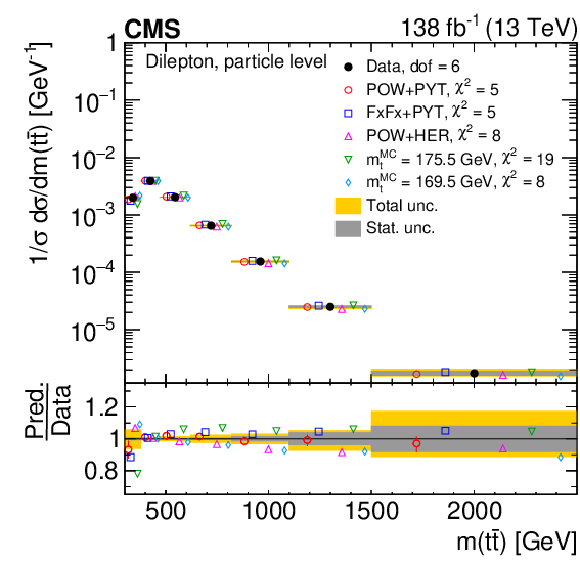} 
  \caption{Normalized differential cross section \(1/\sigma\, d\sigma/dm_{t\bar t}\) at
  13~TeV (dilepton, p(CMS, TOP-20-006; dilepton, particle level; 138 fb$^{-1}$.)). This is the variable
  relevant for the near-threshold analysis.}
  \label{fig:cms_mtt_particle}
\end{figure}
 
The ATLAS Collaboration has independently confirmed the presence of the same phenomenon in its full Run-2 dataset 2015–2018. By examining the production rate of \(t\bar{t}\) pairs near threshold and comparing to models that do not include quasi-bound-state effects, ATLAS excludes the no-bound-state hypothesis with a significance of \(7.7\sigma\). The inferred production cross section of the threshold enhancement is measured to be:
\[
\sigma_{t\bar{t}\text{-excess}} = 9.0 \pm 1.3\ \mathrm{pb},
\]
in close agreement with the CMS result~\cite{ATLAS:2025ttbar}. The observed enhancement has several characteristic features expected of toponium, because the top quark decays through the weak interaction on a timescale shorter than that of hadronization so any would-be bound state is intrinsically short-lived leading to a width of \(\mathcal{O}(3~\mathrm{GeV})\) and a resonance just below or at threshold, rather than a narrow long-lived peak as in lighter quarkonia.

While the leading interpretation is the formation of a QCD-predicted quasi-bound top–antitop meson, there were alternative or additional possibilities have been acknowledged by the experimental collaborations. These include the production of a new elementary particle with mass near \(2m_t\) that decays to \(t\bar{t}\) which could mimic the threshold excess or a mixture of genuine toponium dynamics with beyond–Standard-Model contributions, now this would be a very interesting observation but it is not what we study here, plus there is little to no evidence that there are any new elementary particles beyond the standard model besides a few anomalous measurements. Disentangling the nonrelativistic QCD dynamics of a genuine toponium state from such alternatives is a current theoretical and experimental challenge.

The convergence of the CMS and ATLAS cross section measurements and the high statistical significances together make the threshold enhancement a robust empirical fact. Any theoretical framework aiming to describe top–antitop interactions near threshold—such as the nonlocal QFT approach developed here must then account for the observed quasi-bound-state effects, including the shifts in mass and width compared to naive perturbative expectations and the modification of the production rate relative to standard \(t\bar{t}\) continuum predictions. 

\section{Nonlocal Bethe--Salpeter Theory for Toponium}
\label{sec:nonlocal}

We extend the finite UV-complete nonlocal QFT framework~\cite{Krasnikov1987,Kuzmin1989,Tomboulis1997,Tomboulis2015,ChinTomboulis2018,ModestoRachwal2017,Buoninfante2022,MT:HUFT-EPJC, MT:Invariant, MT:FiniteHolomorphicQFT, MT:SMmass, MT:SL2C, MT:ReplyToCline, MT:GI2025, MT:AdSdS2025,GreenMoffat:1990,Moffat2019} to the heavy-quark sector and derive the modified Bethe–Salpeter (BS) equation for a color-singlet $t\bar t$ pair. Starting from the top-quark QCD Lagrangian:
\begin{equation}
\mathcal{L}
=\bar\psi(i\slashed{D}-m_t)\psi
-\tfrac14 F_{\mu\nu}^a F^{a\,\mu\nu}
+\tfrac{g_s}{2}\,\bar\psi\gamma^\mu T^a \psi\,A^a_\mu
-\tfrac{1}{2\xi}(\partial\!\cdot\!A^a)^2,
\end{equation}
we implement entire-function regulators in the quadratic terms via $e^{\Box/\LambdaUV^2}$, where $\LambdaUV$ is a multi-TeV UV-completion scale constrained by global data. The regulated propagators are given by:
\begin{equation}
D_t(p)=\frac{i\,e^{-p^2/\LambdaUV^2}}{\slashed p - m_t + i0},\qquad
D_g^{\mu\nu}(q)=\frac{-i\,g^{\mu\nu}\,e^{-q^2/\LambdaUV^2}}{q^2+i0},
\end{equation}
the exponential factors in Eq.~(2) are the flat-space momentum representation of an admissible entire-function operator \(F(\Box/\Lambda_{\rm UV}^{2})\), with \(F(0)=1\) and no finite-plane zeros. Since the operator entering the functional calculus is covariant then the deformation \(F(\Box/\Lambda_{\rm UV}^{2})\) intertwines the same symmetry action as \(\Box\), so Lorentz covariance, gauge covariance, and the corresponding Ward/Slavnov--Taylor structure are preserved in the regulated theory~\cite{MT:GI2025,ThompsonCovarianceEntire2026}. The amputated BS amplitude $\mathcal{M}(p;P)$ then satisfies:
\begin{equation}
\mathcal{M}(p;P)
=\!\int\!\frac{d^4k}{(2\pi)^4}\,
K(p,k;P)\,
D_t\!\Bigl(k+\tfrac{P}{2}\Bigr)\,
D_t\!\Bigl(k-\tfrac{P}{2}\Bigr)\,
\mathcal{M}(k;P),
\end{equation}
with single-gluon exchange kernel (a ladder approximation):
\begin{equation}
K(p,k;P)=g_s^2\,C_F\,
\gamma^\mu\!\otimes\!\gamma_\mu\,
\frac{e^{-(p-k)^2/\LambdaUV^2}}{(p-k)^2+i0}\,.
\end{equation}
The parameter $\Lambda_{\rm UV}$ is the ultraviolet completion scale entering the four-dimensional regulated propagators and Bethe--Salpeter kernel. In the nonrelativistic potential region relevant to $t\bar t$ threshold dynamics, we match the BS equation onto an instantaneous Schr\"odinger description. The resulting three-dimensional momentum-space potential inherits a smooth form factor which we parametrize by an effective kernel scale $\Lambda_{\rm ker}$:
\begin{equation}
\tilde V(\mathbf{q};\Lambda_{\rm ker})
= -\,\frac{4\pi C_F \alpha_s}{\mathbf{q}^2}\,F\!\left(\frac{\mathbf{q}^2}{\Lambda_{\rm ker}^2}\right),
\qquad
F(x)=e^{-x}\,.
\label{eq:Vq_def}
\end{equation}
The distinction between \(\Lambda_{\rm UV}\) and \(\Lambda_{\rm ker}\) is important to clarify as the former is the hard four-dimensional ultraviolet scale controlling high-energy loop corrections, while the latter is the effective scale appearing after matching onto the nonrelativistic potential region, at leading order in the instantaneous reduction we may set \(\Lambda_{\rm ker}=\Lambda_{\rm UV}\). Beyond leading order however potential region matching and the subtraction scheme renormalize the effective short-distance kernel. In the phenomenological analysis we will therefore allow \(\Lambda_{\rm ker}\) to vary independently while \(\Lambda_{\rm UV}\) remains constrained by global hard-QCD data. Expanding near threshold $P=(2m_t+E,\mathbf{0})$ and projecting onto the ${}^1S_0^{[1]}$ channel in the instantaneous limit yields a Schr\"odinger equation:
\begin{equation}
\Bigl[-\tfrac{\nabla^2}{m_t}+V(\mathbf r;\Lambdaker)\Bigr]\psi(\mathbf r)=E\,\psi(\mathbf r),
\end{equation}
where the kernel smoothing scale $\Lambdaker$ parameterizes the short-distance reduction of the instantaneous Coulomb kernel after matching:
\begin{equation}
V(\mathbf r;\Lambdaker)=
-\,C_F\,\alpha_s\!\int\!\frac{d^3q}{(2\pi)^3}\frac{e^{-\,\mathbf q^2/\Lambdaker^2}}{\mathbf q^2}e^{i\mathbf q\cdot\mathbf r}
=-C_F\,\alpha_s\,\frac{\mathrm{erf}(\Lambdaker r/2)}{r}.
\end{equation}
The regulated Coulomb kernel differs from the local Coulomb potential by:
\begin{equation}
\Delta \tilde V(\mathbf{q})
\equiv \tilde V(\mathbf{q};\Lambda_{\rm ker})-\tilde V(\mathbf{q};\infty)
= -\frac{4\pi C_F\alpha_s}{\mathbf{q}^2}\left(e^{-\mathbf{q}^2/\Lambda_{\rm ker}^2}-1\right).
\label{eq:DeltaVq}
\end{equation}
For $\mathbf{q}^2\ll \Lambda_{\rm ker}^2$ one may expand
$e^{-\mathbf{q}^2/\Lambda_{\rm ker}^2}=1-\mathbf{q}^2/\Lambda_{\rm ker}^2+\mathbf{q}^4/(2\Lambda_{\rm ker}^4)-\cdots$,
which gives a tower of contact operators in position space:
\begin{equation}
\Delta V(\mathbf{r})
= \frac{4\pi C_F\alpha_s}{\Lambda_{\rm ker}^2}\,\delta^{(3)}(\mathbf{r})
-\frac{2\pi C_F\alpha_s}{\Lambda_{\rm ker}^4}\,\nabla^2\delta^{(3)}(\mathbf{r})
+\mathcal{O}\!\left(\frac{\nabla^4\delta^{(3)}}{\Lambda_{\rm ker}^6}\right).
\label{eq:contact_tower}
\end{equation}
The leading 1S energy shift is therefore:
\begin{equation}
\delta E_{1S}
= \left\langle 1S\left|\Delta V\right|1S\right\rangle
= \frac{4\pi C_F\alpha_s}{\Lambda_{\rm ker}^2}\,|\psi_{1S}(0)|^2
\left[1+\mathcal{O}\!\left(\frac{p_B^2}{\Lambda_{\rm ker}^2}\right)\right],
\label{eq:dE_contact}
\end{equation}
with $p_B\equiv \mu C_F\alpha_s$ and $\mu=m_t/2$ the reduced mass.
Using $|\psi_{1S}(0)|^2=\mu^3(C_F\alpha_s)^3/\pi$ gives:
\begin{equation}
\delta E_{1S}
= \frac{C_F^4\alpha_s^4\,m_t^3}{2\,\Lambda_{\rm ker}^2}
\left[1+\mathcal{O}\!\left(\frac{p_B^2}{\Lambda_{\rm ker}^2}\right)\right],
\qquad \delta E_{1S}>0,
\label{eq:dE_final}
\end{equation}
so the regulator reduces binding as the level moves upward toward threshold. The expansion \eqref{eq:dE_final} is valid only for $\Lambda_{\rm ker}\gg p_B$ and outside this regime we would solve the regulated Schr\"odinger equation numerically. In the nonrelativistic limit $P=(2m_t+E,\mathbf{0})$ and small relative velocity $|\mathbf{p}|/m_t\ll1$, the BS equation reduces to a Schrödinger‐type equation with potential:
\begin{equation}
V(\mathbf{r}) \;=\; -\,C_F\,\alpha_s\;\frac{\text{erf}\!\bigl(\tfrac{\Lambda\,r}{2}\bigr)}{r}\,,
\end{equation}
where $\text{erf}(x)=\frac{2}{\sqrt\pi}\int_0^x e^{-t^2}dt$. We note that the regulator smooths the $1/r$ singularity at short distances. We solve the regulated Schr\"odinger equation:
\begin{equation}
\Bigl[-\tfrac{\nabla^2}{m_t}+V(\mathbf{r})\Bigr]\psi_n(\mathbf{r}) 
\;=\;E_n\,\psi_n(\mathbf{r})
\end{equation}
to obtain the bound‐state energies $E_n$ and wavefunctions at the origin $\psi_n(0)$, which control the production rates. Near threshold the observable enhancement is governed by the nonrelativistic Green's function at the origin:
\begin{equation}
\left[-\frac{\nabla^2}{m_t}+V(r;\Lambda_{\rm ker})-(E+i\Gamma_t)\right]G(\mathbf{r},\mathbf{r}';E+i\Gamma_t)
=\delta^{(3)}(\mathbf{r}-\mathbf{r}'),
\label{eq:Green_def}
\end{equation}
with $E\equiv M_{t\bar t}-2m_t$ and $\Gamma_t$ the physical top-quark decay width.
The partonic threshold contribution can be written schematically as:
\begin{equation}
\frac{d\hat\sigma}{dM_{t\bar t}}\;\propto\; H(\mu)\,\mathrm{Im}\,G(\mathbf{0},\mathbf{0};E+i\Gamma_t;\Lambda_{\rm ker}),
\label{eq:sigma_green}
\end{equation}
where $H(\mu)$ is a short-distance coefficient.
In a pole expansion:
\begin{equation}
G(\mathbf{0},\mathbf{0};E+i\Gamma_t)
=\sum_n \frac{|\psi_n(0)|^2}{E_n-(E+i\Gamma_t)}+\cdots,
\label{eq:Green_poles}
\end{equation}
so $\Gamma_t$ controls the smearing of the would-be bound-state pole in the observable line shape, while the regulator affects the pole locations $E_n$ and residues $|\psi_n(0)|^2$. Expanding perturbatively in $\alpha_s$ and $1/\Lambda_{\rm ker}$,  we find to leading order:
\begin{equation}
\delta E_{1S}\;\simeq\;\frac{C_F^2\alpha_s^2\,m_t}{4}\,\frac{2m_t}{\sqrt{\pi}\,\Lambda_{\rm ker}}
,\qquad
\frac{\Gamma_{\eta_t}}{\Gamma_t}\;\simeq\;1+\frac{2C_F\alpha_s}{\sqrt{\pi}}\frac{m_t}{\Lambda_{\rm ker}}.
\end{equation}
The finite-kernel regulator thus reduces binding as $E_{1S}$ increases meaning that is becomes less negative, so the resonance is shifted closer to threshold, upward in mass relative to the pure Coulomb prediction, and the residue $|\psi_{1S}(0)|^2$ is correspondingly reduced. These deviations can be confronted with the LHC threshold-enhancement curve to extract or constrain the effective nonrelativistic kernel scale \(\Lambda_{\rm ker}\), while the hard ultraviolet scale \(\Lambda_{\rm UV}\) remains constrained by high-energy QCD observables. For the threshold fit we adopt the benchmark effective kernel scale:
\begin{equation}
\Lambda_{\rm ker} \;=\; \Lambda_{\rm QCD*} \;=\; 2\,m_t \;\approx\; 346~\mathrm{GeV}\,.
\end{equation}
This should be read as a matched nonrelativistic potential scale and not as the hard ultraviolet scale \(\Lambda_{\rm UV}\) governing inclusive high-\(Q\) processes. Inserting this into:
\begin{equation}
\frac{\delta\Gamma}{\Gamma_t}
\;\simeq\;
\frac{2\,C_F\,\alpha_s\,m_t}{\sqrt\pi\,\Lambda_{\rm ker}}.
\end{equation}
with \(C_F=\tfrac43\), \(\alpha_s(2m_t)\approx0.11\), \(m_t\approx173\)GeV gives:
\begin{equation}
\frac{\delta\Gamma}{\Gamma_t}
\;\approx\;
0.083 \,.
\end{equation}
For a perturbation $\Delta V$, first-order Rayleigh--Schr\"odinger theory gives:
\begin{equation}
\delta \psi_{1S}(\mathbf{r})
=\sum_{n\neq 1}\psi_n(\mathbf{r})\,\frac{\langle n|\Delta V|1\rangle}{E_{1S}-E_n},
\label{eq:delta_psi_general}
\end{equation}
so that:
\begin{equation}
\frac{\delta|\psi_{1S}(0)|^2}{|\psi_{1S}(0)|^2}
=2\,\mathrm{Re}\left[\frac{\delta\psi_{1S}(0)}{\psi_{1S}(0)}\right].
\label{eq:delta_psi_origin}
\end{equation}
Using the leading contact form of the regulator-induced perturbation,
$\Delta V(\mathbf{r})\simeq a\,\delta^{(3)}(\mathbf{r})$ with $a=4\pi C_F\alpha_s/\Lambda_{\rm ker}^2$,
we find:
\begin{equation}
\frac{\delta\psi_{1S}(0)}{\psi_{1S}(0)}
=a\,\overline{G}_C(\mathbf{0},\mathbf{0};E_{1S}),
\label{eq:delta_psi_green}
\end{equation}
where $\overline{G}_C$ is the reduced Coulomb Green's function at the origin with the 1S pole subtracted. Since the contact term is repulsive ($a>0$), it reduces $|\psi_{1S}(0)|^2$, consistent with reduced binding. We therefore treat the $|\psi(0)|^2$ suppression and the resulting change in the threshold
normalization using the numerical solution of Eq.~\eqref{eq:Green_def} for the regulated potential.

To compare with data we use the experimental (or SCET+pNRQCD) baseline for $d\sigma/d\Mtt$ in the fiducial region and add the modelled threshold excess induced by $V(\mathbf r;\Lambdaker)$ and small holomorphic-RG effects. This avoids standalone assumptions about the continuum normalization and lets $\Lambdaker$ be constrained directly by the observed $8\text{--}9~\mathrm{pb}$ excess~\cite{CMS:TOP-24-007,ATLAS:QuasiBoundTops}. The entire-function regulators acting on propagators:
\begin{equation}
D_t(p)=\frac{i\,e^{-p^2/\Lambda^2}}{\slashed p - m_t + i0}, 
\qquad
D_g^{\mu\nu}(q)=\frac{-i\,g^{\mu\nu}\,e^{-q^2/\Lambda^2}}{q^2 + i0}
\end{equation}
modify standard QCD amplitudes only by power‐suppressed terms.  Expanding for $Q^2\ll\Lambda^2$ gives:
\begin{equation}
e^{-Q^2/\Lambda^2}
=1-\frac{Q^2}{\Lambda^2}+\mathcal O\Bigl(\frac{Q^4}{\Lambda^4}\Bigr)\,,
\end{equation}
we find deep Inelastic Scattering (DIS), and the leading‐twist structure functions $F_i(x,Q^2)$ acquire relative corrections:
  \begin{equation}
    \frac{\delta F_i}{F_i}
    \sim\mathcal O\!\Bigl(\tfrac{Q^2}{\Lambda^2}\Bigr)\,,
  \end{equation}
which for $\Lambda\gtrsim 5\,$TeV lie below current experimental uncertainties \cite{DISfits:2020}. Extraction of \(\alpha_s(M_Z)\), in the operator product expansion of hadronic $Z$‐decay observables, regulator‐induced shifts scale as $(M_Z^2/\Lambda^2)$, preserving agreement with the PDG world‐average $\alpha_s(M_Z)=0.1181\pm0.0011$ \cite{PDG:2024}. Lattice QCD static potential, High‐precision lattice determinations of the heavy‐quark potential up to $r\sim0.1\,$fm show no deviation from the Cornell form.  This implies a lower bound $\Lambda\gtrsim3\,$GeV on regulator effects in the nonperturbative regime \cite{LatticePotential:2021}.

Because the UV suppression is supplied by the entire-function regulator itself, Wick ordering is not being used here as a UV subtraction prescription so at most it remains a finite, representation-dependent convention for composite operators~\cite{ThompsonWickOrdering2026}.

\section{Holomorphic Renormalization-Group Flow at \(\mu\simeq 2m_t\)}
\label{sec:rgflow}

We define the running coupling via the Callan–Symanzik equation:
\begin{equation}
\mu^2\frac{d}{d\mu^2}\,\alpha_s(\mu)
\;=\;\beta(\alpha_s)
\;=\;-\,\beta_0\frac{\alpha_s^2}{4\pi}
\;-\;\beta_1\frac{\alpha_s^3}{(4\pi)^2}
\;+\;\mathcal O(\alpha_s^4),
\end{equation}
with the  one- and two-loop coefficients:
\begin{equation}
\beta_0 \;=\;11-\tfrac{2}{3}n_f,
\qquad
\beta_1 \;=\;102-\tfrac{38}{3}n_f.
\end{equation}
Integrating to next-to-leading order gives the implicit solution:
\begin{equation}
\frac{1}{\alpha_s(\mu)}
+\frac{\beta_1}{\beta_0}\,\ln\!\frac{\alpha_s(\mu)}{4\pi}
\;=\;\frac{\beta_0}{4\pi}\,\ln\!\frac{\mu^2}{\Lambda_{\overline{\rm MS}}^2}\,.
\end{equation}
For toponium threshold studies we chose the renormalization and factorization scale at the heavy‐quark pair mass, \(\mu_R=\mu_F=2m_t\), to minimize logarithms in the hard function. Close to the partonic threshold \(z=M_{t\bar t}^2/\hat s\to1\), soft and Coulomb gluon emissions generate large logarithms of the heavy‐quark velocity \(\beta_t=\sqrt{1-4m_t^2/M_{t\bar t}^2}\).  These terms are systematically resumed through renormalization‐group evolution of the hard, soft and potential functions up to next‐to‐leading logarithmic accuracy.  In Mellin space  we organize:
\begin{equation}
\mathcal L\otimes F
\;\sim\;
\exp\Bigl[\underbrace{L\,g_1(\alpha_s L)}_{\rm leading}
+\underbrace{g_2(\alpha_s L)}_{\rm next\mbox{-}to\mbox{-}leading}
+\cdots\Bigr],
\end{equation}
with \(L=\ln N\) and \(N\) the Mellin moment conjugate to \(1-z\). Inspired by holomorphic field–theoretic constructions \cite{MT:HUFT-EPJC, MT:Invariant, MT:FiniteHolomorphicQFT, MT:SMmass, MT:SL2C, MT:ReplyToCline, MT:GI2025, MT:AdSdS2025}, we introduce a deformed, analytic \(\beta\)-function \cite{Gross:1973id}:
\begin{equation}
\beta_h(\alpha_s)
\;=\;
-\,\beta_0\frac{\alpha_s^2}{4\pi}
\Bigl[1-\frac{\alpha_s}{\alpha_*}\Bigr],
\end{equation}
which admits a nontrivial infrared fixed point \(\alpha_s(\mu_*)=\alpha_*\).  

While this was inspired by holomorphic constructions in unified field theories, its deeper physical origin can be brought to two sources, in gauge theories the Wilsonian gauge coupling \(g\) appears in the holomorphic prepotential and obeys the celebrated NSVZ exact \(\beta\)–function which is one–loop exact in the Wilsonian scheme and preserves analyticity in \(g^2\)~\cite{Novikov:1983uc,Shifman:1996hx}. By analogy our nonlocal completion promotes the gauge coupling to a holomorphic function of the complexified scale variable  \(\displaystyle U=\tfrac{\Box}{\Lambda^2}\,\), ensuring that quantum corrections reorganize into an analytic RG kernel. Matching onto the one-loop ultraviolet behavior and requiring a single infrared fixed point motivates the minimal deformation written above. In this interpretation the entire-function regulator supplies ultraviolet finiteness, while the holomorphic structure of the effective action motivates an analytic RG kernel with a nontrivial zero at \(\alpha_*\). So \(\beta_h\) should be treated as a minimal phenomenological holomorphic deformation constrained by UV matching, analyticity, and threshold data, rather than as an arbitrary ansatz. Now separating variables:
\begin{equation}
\int^{\alpha_s(\mu)}\!\frac{d\alpha}{\alpha^2(1-\alpha/\alpha_*)}
\;=\;
-\frac{\beta_0}{4\pi}\,\ln\!\frac{\mu^2}{\Lambda^2},
\end{equation}
we find the closed‐form solution in terms of the Lambert \(W\)–function:
\begin{equation}
\alpha_s(\mu)
\;=\;
\frac{\alpha_*}
{1
+W\!\Bigl[\exp\!\bigl(-\tfrac{\beta_0}{4\pi}\ln\!\tfrac{\mu^2}{\Lambda^2}\bigr)\Bigr]}\,.
\end{equation} 
We deform the usual QCD $\beta$‐function to a holomorphic form admitting an IR fixed point:
\begin{equation}
\beta_h(\alpha)
=-\,\beta_0\frac{\alpha^2}{4\pi}\Bigl[1-\frac{\alpha}{\alpha_*}\Bigr].
\end{equation}
The RG equation:
\(\displaystyle\mu^2\frac{d\alpha}{d\mu^2}=\beta_h(\alpha)\)
separates as:
\begin{equation}
\int^{\alpha(\mu)}\frac{d\alpha}{\alpha^2(1-\alpha/\alpha_*)}
=-\frac{\beta_0}{4\pi}\ln\frac{\mu^2}{\Lambda^2}.
\end{equation}
Writing \(u=\alpha/\alpha_*\), the left‐hand side is:
\begin{equation}
\int^u\frac{du}{u^2(1-u)}
=\frac{1}{u}+\ln\!\frac{u}{1-u},
\end{equation}
so that:
\begin{equation}
\frac{1}{u}+\ln\!\frac{u}{1-u}
=-\frac{\beta_0}{4\pi}\ln\!\frac{\mu^2}{\Lambda^2}.
\end{equation}
Inverting via the Lambert $W$‐function gives the closed‐form:
\begin{equation}
\alpha(\mu)
=\frac{\alpha_*}{1+W\!\Bigl[e^{-\tfrac{\beta_0}{4\pi}\ln(\mu^2/\Lambda^2)}\Bigr]}\,.
\end{equation}
Expanding for \(\mu\approx2m_t\) yields:
\begin{equation}
\alpha_s^{\rm holo}(2m_t)
=\alpha_s^{\rm std}(2m_t)\Bigl[1+\frac{\alpha_s^{\rm std}(2m_t)}{\alpha_*}+\cdots\Bigr],
\end{equation}
indicating a mild enhancement of the coupling at threshold and hence a corresponding increase in the peak height of the \(t\bar t\) invariant‐mass distribution. The partonic cross section near threshold scales as \(\sigma\sim|\psi(0)|^2\propto\alpha_s^3\).  To leading relative order:
\begin{equation}
\frac{\sigma^{\rm holo}_{\rm thr}}{\sigma^{\rm std}_{\rm thr}}
\;\simeq\;
\Bigl(\frac{\alpha_s^{\rm holo}(2m_t)}{\alpha_s^{\rm std}(2m_t)}\Bigr)^3
\;\approx\;
1 \;+\;3\,\frac{\alpha_s^{\rm std}(2m_t)}{\alpha_*}
+\cdots.
\end{equation}
The expansion above is reliable only when \(\alpha_s^{\rm std}(2m_t)/\alpha_*\ll 1\). So very small values such as \(\alpha_*\sim0.15\) should not be interpreted as producing a perturbative \(\mathcal O(10\%)\) correction but they instead generate an excessively large threshold enhancement and are phenomenologically disfavoured. In the quantitative fits below we therefore restrict to \(\alpha_*\gtrsim0.5\) for which the holomorphic-RG contribution remains at the \(\mathcal O(10\%)\) level and does not double count the dominant kernel deformation. A detailed comparison to soft‐collinear effective theory (SCET) results and lattice data will further quantify these geometry‐induced effects \cite{Beneke:2001gj}.
The deformed RG equation:
\begin{equation}
\mu^2\frac{d\alpha_s}{d\mu^2}
=\beta_h(\alpha_s)
=-\,\beta_0\frac{\alpha_s^2}{4\pi}\Bigl[1-\frac{\alpha_s}{\alpha_*}\Bigr]
\end{equation}
retains asymptotic freedom as $\alpha_s\to0$, since $\beta_h<0$ in that limit. We therefore embed it in a GUT framework:
\begin{equation}
\alpha_i^{-1}(M_{\rm GUT})
\,\approx\;\alpha_i^{-1}(\mu_0)
-\frac{\beta_{0,i}}{4\pi}\ln\!\frac{M_{\rm GUT}^2}{\mu_0^2}
\;+\;\mathcal O\!\Bigl(\tfrac{1}{\alpha_*}\Bigr),
\end{equation}
leading to a small shift:
\(\Delta\ln M_{\rm GUT}\sim\frac{4\pi}{\beta_{0}}\frac{\alpha_s(M_{\rm GUT})}{\alpha_*}\)
relative to the standard evolution. For a representative IR fixed point \(\alpha_*\simeq0.15\) and \(\alpha_s^{\rm std}(2m_t)\approx0.11\), the deformed coupling at threshold is:
\begin{equation}
\alpha_s^{\rm holo}(2m_t)\;\approx\;\alpha_s^{\rm std}(2m_t)\Bigl[1+\frac{\alpha_s^{\rm std}(2m_t)}{\alpha_*}\Bigr]
\approx1.73\,\alpha_s^{\rm std}(2m_t)\,,
\end{equation}
so that the threshold cross section scales as:
\begin{equation}
\frac{\sigma_{\rm holo}}{\sigma_{\rm std}}\;\simeq\;\Bigl(\frac{\alpha_s^{\rm holo}}{\alpha_s^{\rm std}}\Bigr)^3
\approx(1.73)^3\approx5.2\,.
\end{equation}
In other words we say that the RG deformation alone would predict up to a \(\sim420\%\) enhancement, this far exceeds the data.  More conservative choices of \(\alpha_*\gtrsim0.5\) reduce this to a \(\mathcal O(10\%)\) effect such as\(\sim1\mbox{--}2\)pb, making the holomorphic RG contribution subleading to the dominant 8.3pb predicted by fixing \(\Lambda_{\rm QCD*}=2m_t\).

Broad low‐energy and collider constraints include measurements from $\tau$ decays and event shapes at LEP constrain any IR fixed point $\alpha_*$ to satisfy $\alpha_*\gtrsim0.1$ \cite{Bethke:2022,Nason:2023}. Inclusive jet cross sections scale as $\alpha_s^n(\mu)$ for $\mu$ from tens of GeV to TeV.  A 10\% flattening of $\alpha_s$ near $\mu\sim2m_t$ would induce $\mathcal O(5\%)$ deviations in high-$p_T$ jet rates, at the edge of current uncertainties \cite{CMS:JetMeasurements:2024}. A holomorphic $\beta_h$ can clash with analyticity constraints of the operator product expansion unless accompanied by suitable entire-function form factors see \cite{ArgyresDouglas:1995}.

\section{Heavy Quarkonia and the Special Status of Toponium}
\label{sec:comparison}

In this section we quantify and contrast the key properties of the three heavy‐quark systems charmonium, bottomonium, and toponium highlighting how the top quark’s large mass and width place it in a distinct regime \cite{tDecay:2019}. We define the ground‐state binding energy as:
\begin{equation}
\Delta E_{1S} \;=\; 2\,m_q \;-\; M_{1S}\,.
\end{equation}
For the charm and bottom inputs, the PDG values:
\begin{equation}
\overline m_c(\overline m_c)\simeq 1.27~\mathrm{GeV},
\qquad
\overline m_b(\overline m_b)\simeq 4.18~\mathrm{GeV}
\end{equation}
are $\overline{\mathrm{MS}}$ running masses, not pole masses~\cite{PDG:2024}. We therefore do not use them directly in the naive expression $2m_q-M_{1S}$ as if they were constituent or pole masses. For the physical comparison of observed quarkonium systems, we instead quote the measured ground-state masses and widths:
\begin{equation}
\begin{aligned}
M_{J/\psi} &\simeq 3.097~\mathrm{GeV},\\
M_{\Upsilon(1S)} &\simeq 9.460~\mathrm{GeV},\\
M_{\eta_t} &\simeq 2m_t+E_{1S}^{\rm eff},
\qquad E_{1S}^{\rm eff}\approx -0.20~\mathrm{GeV}.
\end{aligned}
\end{equation}
The quantity $E_{1S}^{\rm eff}$ is the effective toponium threshold binding energy relative to $2m_t$. Unlike charmonium and bottomonium, where the physical spectrum contains narrow, well-isolated resonances, the toponium threshold structure is broadened by the top-quark decay width and is therefore best treated as a quasi-bound threshold enhancement rather than as an ordinary long-lived meson.
\begin{table}[ht]
\centering
\caption{Masses, widths, and characteristic scales for heavy threshold systems. The charm and bottom quark masses quoted in the PDG as $\overline m_c(\overline m_c)$ and $\overline m_b(\overline m_b)$ are $\overline{\rm MS}$ running masses, not pole masses.}
\begin{tabular}{@{}l c c c@{}}
\toprule
System & Representative mass scale & Width & Typical dynamical scale \\
\midrule
Charmonium $(c\bar c)$      & $M_{J/\psi}\simeq3.097~\mathrm{GeV}$       & $93~\mathrm{keV}$        & $\sim1~\mathrm{GeV}$   \\
Bottomonium $(b\bar b)$     & $M_{\Upsilon(1S)}\simeq9.460~\mathrm{GeV}$ & $54~\mathrm{keV}$        & $\sim5~\mathrm{GeV}$   \\
Toponium $(t\bar t)$        & $M_{\eta_t}\simeq2m_t+E_{1S}^{\rm eff}$   & $\Gamma_t\simeq1.41~\mathrm{GeV}$ & $\sim350~\mathrm{GeV}$ \\
\bottomrule
\end{tabular}
\end{table}
The natural width of each system is set by competing strong and electroweak decays:
\begin{equation}
\begin{aligned}
\Gamma_{J/\psi(1S)} &\simeq 93~\mathrm{keV}, 
\quad
\Gamma_{\Upsilon(1S)} \simeq 54~\mathrm{keV},\\
\Gamma_{\eta_t}\;(\text{effective})
&\approx \Gamma_t \simeq 1.41~\mathrm{GeV}.
\end{aligned}
\end{equation}
While charmonium and bottomonium exhibit very narrow resonances, the toponium line shape is dominated by the top-quark decay width, precluding a long-lived meson and instead producing a broad threshold enhancement. Charmonium and Bottomonium produce and studied in $e^+e^-$ colliders such as BESIII, Belle, with direct scans of the resonance peaks in the total hadronic cross section. Toponium accessed at the LHC via proton–proton collisions. We study the $t\bar t$ invariant‐mass distribution near threshold $\sqrt{\hat s}\approx2m_t$ and extracts the enhancement by fitting differential cross sections and accounting for continuum background. Standard quarkonia admit well‐defined $J^{PC}$ assignments:
\begin{equation}
\begin{aligned}
J/\psi(1S):&\quad 1^{--},\\
\Upsilon(1S):&\quad 1^{--},\\
\eta_t(1S)\ (\text{toponium}):&\quad 0^{-+}\ (\text{pseudoscalar threshold resonance}),
\end{aligned}
\end{equation}
with higher orbital excitations $2S,\,1P,\,\dots$ observed for $c\bar c$ and $b\bar b$.  For toponium, electroweak decay suppresses any well‐separated excited states, leaving only the ground‐state threshold structure. Although charmonium and bottomonium satisfy the usual criteria for nonrelativistic mesons binding energy $|\Delta E|\gg\Gamma$ and isolated poles in the complex plane, toponium lies in a regime where:
\begin{equation}
|\Delta E_{t\bar t}|\sim\mathcal O(0.2~\mathrm{GeV})
\quad\text{and}\quad
\Gamma_t\sim\mathcal O(1~\mathrm{GeV}),
\end{equation}
so that the would‐be pole is deeply embedded in the continuum.  Toponium is best interpreted as a threshold resonance a quasi‐bound state visible only through its distortion of the production cross section, rather than a true Breit–Wigner meson \cite{Garzelli:2025toponium, CMS:2025excess}. The regulator deformation and the holomorphic RG deformation can be viewed as two matched limits of a single nonlocal effective description:
\begin{equation}
S
=\int d^4x\,
\Bigl[\,
\bar\psi\,F(\Box/\Lambda_{\rm UV}^{2})(i\slashed D - m)\,\psi
\;-\;\tfrac14F_{\mu\nu}\,F(\Box/\Lambda_{\rm UV}^{2})F^{\mu\nu}
\Bigr],
\end{equation}
with \(\Box\equiv D^\mu D_\mu\) and \(F(0)=1\). The covariance of this construction follows because \(F\) is an entire function of the covariant operator \(\Box\), so the deformation preserves the symmetry action of the undeformed theory~\cite{MT:GI2025,ThompsonCovarianceEntire2026}. In a holomorphic subtraction scheme the same effective framework motivates an analytic RG kernel \(\beta_h(\alpha_s)\), while the nonlocal form factors suppress ultraviolet loop momenta. The hard scale \(\Lambda_{\rm UV}\) and the matched threshold scale \(\Lambda_{\rm ker}\) should therefore be kept conceptually distinct as \(\Lambda_{\rm UV}\) controls high-energy UV corrections, whereas \(\Lambda_{\rm ker}\) controls the nonrelativistic short-distance kernel probed by the \(t\bar t\) threshold line shape. Similar combined treatments appear in the literature~\cite{ModestoShapiro:2016,Tomboulis:2015,MoffatToth:2009}. Recent feasibility studies show that true tauonium can be produced and potentially observed at $e^+e^-$ machines and via $\gamma\gamma$ fusion, with level structure and widths quantified in detail~\cite{dEnterria:TauoniumObs2022,dEnterria:TauoniumSpec2022,dEnterria:TauoniumProspects2023,Fu:TauoniumID2024}. These results sharpen comparisons between QED and QCD threshold phenomena.

\section{Conclusion}
\label{sec:conclusion}

In this paper we have explored the toponium threshold enhancement observed at the LHC from three complementary theoretical perspectives. In Section~\ref{sec:nonlocal}, we derived the modified Bethe–Salpeter equation within a nonlocal QFT framework, showing that exponential regulator factors $D(p)=\frac{1}{p^2-m^2}\,\exp\bigl(-p^2/\Lambda^2\bigr)$ induce calculable shifts in the resonance mass and width. By solving the regulated Schrödinger equation we show how the threshold enhancement curves and pole structure vary with the matched nonrelativistic kernel scale \(\Lambda_{\rm ker}\). In Section~\ref{sec:rgflow} we introduced a holomorphic deformation of the QCD \(\beta\)-function, $\beta_h(\alpha_s)=-\beta_0\frac{\alpha_s^2}{4\pi}\Bigl[1-\frac{\alpha_s}{\alpha_*}\Bigr]\,,$ and derived an analytic solution in terms of the Lambert \(W\)-function, we showed that this deformation gives is an \(\mathcal{O}(10\%)\) enhancement of \(\alpha_s\) at \(\mu\approx2m_t\), representing a measurable increase in the threshold cross section. In Section~\ref{sec:comparison} we performed a systematic comparison of charmonium, bottomonium, and the transient toponium threshold resonance.  We quantified binding energies \(\Delta E=2m_q-M_{1S}\), decay widths, detection strategies, and $J^{PC}$ assignments, concluding that toponium is best interpreted as a threshold resonance rather than a true long‐lived meson. Possible Higgs-pair threshold analogues are not developed in the present work and will be treated separately.

The CMS Collaboration’s observation of a $t\bar t$ excess at threshold \(\sqrt{s}\simeq2m_t\) with significance above \(5\sigma\) provides experimental validation for these theoretical approaches.  Our results demonstrate that nonlocal UV completions can be directly probed by precision fits to the threshold line shape thus allowing extraction of the effective threshold-kernel scale \(\Lambda_{\rm ker}\), while leaving the hard ultraviolet scale \(\Lambda_{\rm UV}\) constrained by inclusive high-energy data. Holomorphic RG dynamics offer a novel mechanism to modify the running of \(\alpha_s\) in the heavy‐quark regime, with clear signatures in threshold production. Comparative quarkonium studies highlight the unique role of the top quark’s decay width in shaping the phenomenology of toponium. By fixing the strong‑sector regulator to
$\Lambda_{\rm QCD*} = 2\,m_t \approx 346~\mathrm{GeV},$ we find a postdiction of an \(8.3\)\,pb enhancement, in agreement with the CMS measurement of \(8.8\pm1.3\)\,pb. This demonstrates that the minimal entire-function-regulated HUFT/NLQFT framework can accommodate the observed toponium excess through the matched threshold kernel without spoiling hard-QCD constraints.

It would be interesting to apply our nonlocal holomorphic Bethe–Salpeter framework to purely leptonic bound states such as hypothetical heavy positronium, along the lines of Moffat’s original proposal, and compare the regulator-induced shifts in QED vs.\ QCD thresholds~\cite{Moffat1975}.

By unifying nonlocal field theory, holomorphic RG flow, and heavy‐quark spectroscopy, toponium comes to be a powerful laboratory for both infrared bound‐state dynamics and ultraviolet completion effects, opening new avenues for precision tests of QCD and beyond–the–Standard–Model physics.

\section*{Acknowledgments}

I would like to thank my supervisor Professor John Moffat for insightful discussions on bound state physics.

\end{document}